\documentclass[english,aps,manuscript,aps,preprint]{revtex4}
\usepackage[T1]{fontenc}
\usepackage[latin9]{inputenc}
\usepackage{textcomp}
\usepackage{amsmath}
\usepackage{amssymb}
\usepackage{esint}

\makeatletter

\providecommand{\tabularnewline}{\\}

\@ifundefined{textcolor}{}
{%
 \definecolor{BLACK}{gray}{0}
 \definecolor{WHITE}{gray}{1}
 \definecolor{RED}{rgb}{1,0,0}
 \definecolor{GREEN}{rgb}{0,1,0}
 \definecolor{BLUE}{rgb}{0,0,1}
 \definecolor{CYAN}{cmyk}{1,0,0,0}
 \definecolor{MAGENTA}{cmyk}{0,1,0,0}
 \definecolor{YELLOW}{cmyk}{0,0,1,0}
 }

\makeatother

\usepackage{babel}

\begin{document}

\title{One-loop quantum gravity repulsion in the early Universe}

\thanks{Dedicated to Jakub Rembieli\'{n}ski on the occasion of his 65th birthday.}

\author{Bogus\l{}aw Broda}

\email{bobroda@uni.lodz.pl}

\homepage{http://merlin.fic.uni.lodz.pl/kft/people/BBroda}

\affiliation{Department of Theoretical Physics, University of \L{}ód\'{z}, Pomorska
149/153, PL--92-002 \L{}ód\'{z}, Poland}
\begin{abstract}
Perturbative quantum gravity formalism is applied to compute the lowest
order corrections to the classical spatially flat cosmological FLRW
solution (for the radiation). The presented approach is analogous
to the approach applied to compute quantum corrections to the Coulomb
potential in electrodynamics, or rather to the approach applied to
compute quantum corrections to the Schwarzschild solution in gravity.
In the framework of the standard perturbative quantum gravity, it
is shown that the corrections to the classical deceleration, coming
from the one-loop graviton vacuum polarization (self-energy), have
(UV cutoff free) opposite to the classical repulsive properties which
are not negligible in the very early Universe. The repulsive {}``quantum
forces'' resemble those known from loop quantum cosmology.


\textbf{PACS (2010) numbers: }04.60.Gw Covariant and sum-over-histories
quantization; 98.80.Es Observational cosmology (including Hubble constant,
distance scale, cosmological constant, early Universe, etc): 04.60.Bc
Phenomenology of quantum gravity; 04.60.Pp Loop quantum gravity, quantum
geometry, spin foams.
\end{abstract}
\maketitle

\paragraph*{Introduction.}

The aim of our work it to explicitly show the appearance of {}``repulsive
forces'' of quantum origin in the very early Universe. As fundamental
guiding references to our letter we would like to point out the publications
devoted to perturbative calculations of quantum corrections to classical
electromagnetic (the Uehling potential) and gravitational fields.
More precisely, we would mean, for example, the works presenting one-loop
quantum corrections to the Coulomb potential in electrodynamics (see,
e.g.\ §~114 in \cite{QuantumElectrodynamics}), or rather the lowest
order quantum corrections to the Schwarzschild solution in gravity
(see, \cite{QuantumcorrectionstotheSchwarzschildsolution}, and also,
e.g.\  \cite{LeadingquantumcorrectiontotheNewtonianpotential}).
Actually, we apply the method successfully used in the case of the
Schwarzschild solution in \cite{QuantumcorrectionstotheSchwarzschildsolution}
to the spatially flat Friedmann--Lemaître--Robertson--Walker (FLRW)
solution (for the radiation). Fortunately, it appears that the cosmological
FLRW case is only a little bit more complicated than Schwarzschild's
one, and moreover our results conform with the present knowledge.
Namely, the lowest order quantum corrections coming from the fluctuating
graviton vacuum yield {}``repulsive forces'' resembling the well-known
situation in loop quantum cosmology (LQC). The phenomenon is negligible
in our epoch, but it is not so in the very early Universe. Moreover,
it appears, the result is UV cutoff-free, despite the fact that the
cutoff has been primarily imposed (see, e.g.\ \cite{LeadingquantumcorrectiontotheNewtonianpotential}).
One should stress that our derivation is a lowest order approximation---the
graviton vacuum polarization (self-energy) is taken in one-loop approximation,
and the approach assumes the validity of the weak-field regime.

\paragraph*{Quantum corrections.}

Our starting point is a general spatially flat FLRW metric\begin{equation}
ds^{2}=g_{\mu\nu}dx^{\mu}dx^{\nu}=dt^{2}-a^{2}(t)\boldsymbol{dr}^{2},\label{eq:general FLRW metric}\end{equation}
with the cosmic scale factor $a(t)$. To satisfy the condition of
the weakness of the (perturbative) gravitational field $\kappa h_{\mu\nu}$
near our reference time $t=t_{0}$ in the expansion\begin{equation}
g_{\mu\nu}=\eta_{\mu\nu}+\kappa h_{\mu\nu},\label{eq:g=00003D00003D00003D00003D00003D00003D00003D00003D00003D00003D00003D00003D00003D00003D00003D00003D00003D00003D00003D00003D00003D00003D00003Dn+h}\end{equation}
($\kappa=\sqrt{32\pi G_{N}}$, with $G_{N}$---the Newton gravitational
constant), the metric is rescaled such a way that it is exactly Minkowskian
for $t=t_{0}$, i.e.\begin{equation}
a^{2}(t)=1-\kappa h(t),\qquad h(t_{0})=0.\label{eq:metric}\end{equation}
Then\begin{equation}
h_{\mu\nu}(t,\boldsymbol{r})=h(t)\mathcal{I_{\mu\nu}}\quad\mathrm{and}\quad\mathcal{I}_{\mu\nu}\equiv\left(\begin{array}{cc}
0 & 0\\
0 & \delta_{ij}\end{array}\right).\label{eq:h mi ni eq. h I mi ni}\end{equation}
In view of the standard harmonic gauge condition (see, the second
eq.\ in (\ref{eq:our barred metric})) which is to be imposed in
a moment, we perform the following gauge transformation:\begin{equation}
\kappa h_{\mu\nu}\rightarrow\kappa h_{\mu\nu}^{'}=\kappa h_{\mu\nu}+\partial_{\mu}\xi_{\nu}+\partial_{\nu}\xi_{\mu}\quad\textrm{with }\quad\xi_{\mu}\left(t\right)=\left(-\frac{3\kappa}{2}\int_{0}^{t}h(t')\, dt',\,0,\,0,\,0\right).\label{eq:gauge transformation}\end{equation}
Skipping the prime for simplicity, we get\begin{equation}
h_{\mu\nu}(t,\boldsymbol{r})=h(t)\left(\begin{array}{cc}
-3 & 0\\
0 & \delta_{ij}\end{array}\right)\quad\textrm{and }\quad h{}_{\lambda}^{\lambda}(t)=-6h(t),\label{eq:new metric tensor}\end{equation}
where indices are being manipulated with the Minkowski metric $\eta_{\mu\nu}$.
Switching from $h_{\mu\nu}$ to standard ({}``better'') perturbative
gravitational variables, namely to the {}``barred'' field $\bar{h}_{\mu\nu}$
defined by\begin{equation}
\bar{h}_{\mu\nu}\equiv h_{\mu\nu}-\tfrac{1}{2}\eta_{\mu\nu}h_{\lambda}^{\lambda},\label{eq:general barred metric}\end{equation}
we get \begin{equation}
\bar{h}_{\mu\nu}(t,\boldsymbol{r})=-2h(t)\mathcal{I}_{\mu\nu}\quad\textrm{with}\quad\partial^{\mu}\bar{h}_{\mu\nu}=0.\label{eq:our barred metric}\end{equation}
The Fourier transform of $\bar{h}_{\mu\nu}$ is given by\begin{equation}
\tilde{\bar{h}}_{\mu\nu}(p)=-2\tilde{h}(E)\left(2\pi\right)^{3}\delta^{3}(\boldsymbol{p})\mathcal{I}_{\mu\nu}.\label{eq:fourier metric}\end{equation}

To obtain quantum corrections to a classical field (line) we should
supplement the classical line with a vacuum polarization (self-energy)
contribution and a corresponding (full) propagator. Therefore, the
lowest order quantum corrections $\tilde{\bar{h^{\textrm{q}}}}_{\mu\nu}$
to the classical gravitational field $\tilde{\bar{h^{\textrm{c}}}}_{\mu\nu}$
are given, in the momentum representation, by the formula (see, e.g.\ \cite{QuantumcorrectionstotheSchwarzschildsolution},
or §~114 in \cite{QuantumElectrodynamics} for an electrodynamic
version---the Uehling potential)\begin{equation}
\tilde{\bar{h^{\textrm{q}}}}_{\mu\nu}(p)=\left(D\Pi\tilde{\bar{h^{\textrm{c}}}}\right)_{\mu\nu}(p),\label{eq:quantum from classical 1}\end{equation}
where\begin{equation}
D_{\mu\nu}^{\alpha\beta}(p)=\frac{i}{p^{2}}\mathbb{D_{\mu\nu}^{\alpha\beta}}\label{eq:propagator}\end{equation}
is the free graviton propagator in the harmonic gauge with the auxiliary
(constant) tensor $\mathbb{D}$ defined in Eq.(\ref{eq:D,E,P}) below,
and $\Pi_{\mu\nu}^{\alpha\beta}(p)$ is the (one-loop) graviton vacuum
polarization (self-energy) tensor operator. Now, we are defining the
following useful auxiliary tensors:\begin{equation}
\mathbb{D}\equiv\mathbb{E}-2\mathbb{P},\quad\textrm{where}\quad\mathbb{E}_{\mu\nu}^{\alpha\beta}\equiv\tfrac{1}{2}\left(\delta_{\mu}^{\alpha}\delta_{\nu}^{\beta}+\delta_{\nu}^{\alpha}\delta_{\mu}^{\beta}\right)\quad\textrm{and}\quad\mathbb{P}_{\mu\nu}^{\alpha\beta}\equiv\tfrac{1}{4}\eta^{\alpha\beta}\eta_{\mu\nu};\label{eq:D,E,P}\end{equation}
which satisfy the following obvious identities:\begin{equation}
\mathbb{E}^{2}=\mathbb{E},\quad\mathbb{P}^{2}=\mathbb{P},\quad\mathbb{EP}=\mathbb{PE}=\mathbb{P}\quad\textrm{and}\quad\mathbb{D}^{2}=\mathbb{E}.\label{eq:E, P, D identities}\end{equation}
By virtue of the definition (\ref{eq:general barred metric}), we
observe that \begin{equation}
\bar{h}_{\mu\nu}=\left(\mathbb{D}h\right)_{\mu\nu}.\label{eq:h bar eq Dh}\end{equation}
Multiplying Eq.(\ref{eq:quantum from classical 1}) from the left
by $\mathbb{D}$, we obtain (using (\ref{eq:propagator}), (\ref{eq:h bar eq Dh}),
and the last identity in the series (\ref{eq:E, P, D identities}))\begin{equation}
\tilde{h^{\textrm{q}}}_{\mu\nu}(p)=\frac{i}{p^{2}}\left(\Pi\tilde{\bar{h^{\textrm{c}}}}\right)_{\mu\nu}(p).\label{eq:quantum from classical}\end{equation}
Actually, a useful simplification takes place in (\ref{eq:quantum from classical}),
namely,\begin{equation}
\tilde{h^{\textrm{q}}}_{\mu\nu}\left(p\right)=\frac{i}{p^{2}}\left(\Pi'\tilde{\bar{h^{\textrm{c}}}}\right)_{\mu\nu}\left(p\right),\label{eq:quantum from classical prime}\end{equation}
where $\Pi'(p)$ is an {}``essential'' part of the full (in one-loop
approximation) graviton polarization operator $\Pi(p)$. The {}``essential''
part $\Pi'\left(p\right)$ of the full (one-loop) graviton vacuum
polarization operator $\Pi\left(p\right)$ is obtained from $\Pi\left(p\right)$
by skipping all the terms with the momenta $p$ with free indices
(e.g. $\alpha$, $\beta$, $\mu$, or $\nu$). Such a simplification
follows from the gauge freedom the $\tilde{h^{\textrm{q}}}_{\mu\nu}$
enjoys, and from the harmonic gauge condition the ${\tilde{\bar{h^{\textrm{c}}}}}_{\alpha\beta}$
satisfies. In general, by virtue of symmetry of indices, $\Pi\left(p\right)$
consists of 5 (tensor) terms. Each $p_{\mu}$ can be ignored in $\Pi\left(p\right)$
because it only generates a gauge transformation of $\tilde{h}_{\mu\nu}$.
Moreover, since $\bar{\tilde{h}}_{\alpha\beta}$ satisfies the harmonic
gauge condition, the terms with $p^{\alpha}$ in $\Pi\left(p\right)$
are being annihilated. In other words,\begin{equation}
\Pi\left(p\right)=\underbrace{\Pi'\left(p\right)}_{\textrm{2 terms}}+\;\underbrace{\cdots\enskip p\enskip\cdots}_{\textrm{3 skipped terms}}\;.\label{eq:Pi=00003D00003D00003D00003D00003D00003D00003D00003D00003D00003D00003D00003D00003D00003D00003D00003D00003D00003D00003D00003D00003D00003D00003DPi'+...p...}\end{equation}
Since the momenta $p$ in the ellipses posses free indices, they can
be ignored, and only the first two terms with dummy indices ($p^{2}$)
survive ($\Pi'(p)$). Thus,\begin{equation}
\Pi'(p)=\kappa^{2}p^{4}I(p^{2})(2\alpha_{1}\mathbb{E}+4\alpha_{2}\mathbb{P}),\label{eq:Pi'}\end{equation}
where the numerical values of the coefficients $\alpha_{1}$ and $\alpha_{2}$
depend on the kind of the virtual field circulating in the loop, and
the (scalar) standard loop integral $I(p^{2})$ with the UV cutoff
$M$ is asymptotically of the form (see, e.g., Chapt.\ 9.4.2 in \cite{QuarksLeptonsandGaugeFields})\begin{equation}
I(p^{2})=\frac{1}{\left(2\pi\right)^{4}}\intop_{\mathrm{{\scriptscriptstyle UV}\,{\normalcolor {\scriptscriptstyle cutoff}}}={\scriptscriptstyle M}}\frac{d^{4}q}{q^{2}\left(p-q\right)^{2}}=-\frac{i}{\left(4\pi\right)^{2}}\log\left(-\frac{p^{2}}{M^{2}}\right)+\cdots\quad,\label{eq:asymptotic integral}\end{equation}
where the dots mean terms $\mathcal{O}\left(p^{2}/M^{2}\right)$.
A standard way to derive (\ref{eq:asymptotic integral}) consists
in continuing $q_{0}$ to $+iq_{4}$ (which corresponds to Euclidean
formalism, $d^{4}q\rightarrow id^{4}q_{\mathrm{E}})$, exponentiating
the denominator using a (double) proper-time representation for the
propagators, a change of proper-time variables, imposing the UV cutoff
for a new proper time, and continuing back to the Minkowskian momentum.
Thus, we obtain\begin{align}
\tilde{h^{\textrm{q}}}_{\mu\nu}(p) & =\frac{i}{p^{2}}\kappa^{2}p^{4}\left[-\frac{i}{\left(4\pi\right)^{2}}\log\left(-\frac{p^{2}}{M^{2}}\right)\right]\left[-2\tilde{h^{\mathrm{c}}}(E)\left(2\pi\right)^{3}\delta^{3}(\boldsymbol{p})\right]\left[(2\alpha_{1}\mathbb{E}+4\alpha_{2}\mathbb{P})\mathcal{I}\right]_{\mu\nu}\nonumber \\
 & =-2\pi\kappa^{2}E^{2}\log\left|\frac{E}{M}\right|\tilde{h^{\mathrm{c}}}(E)\delta^{3}(\boldsymbol{p})\left(\begin{array}{cc}
-3\alpha_{2} & 0\\
0 & \left(2\alpha_{1}+3\alpha_{2}\right)\delta_{ij}\end{array}\right).\label{eq:quantum h(t)}\end{align}
The unnecessary modulus sign in Eq.(\ref{eq:quantum h(t)}) is only
to remind the fact that there is also an imaginary contribution to
the metric due to creation processes which are ignored in our further
analysis.

\paragraph*{Radiation source.}

Now, we should specify our input classical metric. For definiteness,
we choose the radiation as a source (the early Universe), but it is
not crucial, and assume\begin{equation}
a^{2}(t)=\theta(t)\left(\frac{t}{t_{0}}\right).\label{eq:m-radiation}\end{equation}
Then\begin{equation}
\kappa h^{\mathrm{c}}(t_{0})=0,\qquad\kappa\dot{h^{\mathrm{c}}}(t_{0})=-\frac{1}{t_{0}}\qquad\mathrm{and}\qquad\kappa\ddot{h^{\mathrm{c}}}(t_{0})=0.\label{eq:derivatives m-rad}\end{equation}
By virtue of the definition of the deceleration parameter $q$, expressed
by\begin{equation}
q(t_{0})\equiv-\frac{a\ddot{a}}{\left(\dot{a}\right)^{2}}(t_{0})=1+2\left[1-\kappa h(t_{0})\right]\frac{\kappa\ddot{h}(t_{0})}{\left(\kappa\dot{h}(t_{0})\right)^{2}},\label{eq:general q}\end{equation}
we immediately get the classical result\begin{equation}
q^{\mathrm{c}}(t_{0})=1.\label{eq:qc1}\end{equation}

According to (\ref{eq:metric}) and (\ref{eq:m-radiation}) the Fourier
transform of $h^{\mathrm{c}}(t)$ is\begin{equation}
\tilde{h^{\mathrm{c}}}(E)=\frac{1}{\kappa t_{0}}\left(\frac{1}{E^{2}}+\cdots\right),\label{eq:Fourier(h)}\end{equation}
where the dots mean terms (vanishing in the next formula) proportional
to the Dirac delta and its first derivative. Hence\begin{equation}
\tilde{h^{\mathrm{q}}}_{\mu\nu}(p)=-\frac{2\pi\alpha\kappa}{t_{0}}\log\left|\frac{E}{M}\right|\delta^{3}(\boldsymbol{p})\mathcal{I}_{\mu\nu},\label{eq:Fourier h(p)}\end{equation}
where $\alpha\equiv2\alpha_{1}+3\alpha_{2}$, and performing the gauge
transformation in the spirit of (\ref{eq:gauge transformation}),
we have removed the purely time component of ${h^{\mathrm{q}}}_{\mu\nu}$,
i.e.\ ${h^{\mathrm{q}}}_{00}\rightarrow{h^{\mathrm{q}}}'_{00}=0$.
The inverse Fourier transform yields\begin{equation}
{h^{\mathrm{q}}}_{\mu\nu}(t)=\frac{2\pi^{2}\alpha\kappa}{\left(2\pi\right)^{4}t_{0}}\left(|t|^{-1}+\cdots\right)\mathcal{I}_{\mu\nu},\label{eq:hqmn(t)}\end{equation}
where this time the dots mean a term (vanishing for $t>0$) proportional
to the Dirac delta. Therefore, for $t>0$ we have\begin{equation}
\kappa h^{\mathrm{q}}(t)=\frac{\alpha\kappa^{2}}{8\pi^{2}t_{0}}t^{-1}=-\frac{G}{4\pi t_{0}}t^{-1},\label{eq:khq(t)}\end{equation}
where according to Table \ref{Flo:TABLE I} only the graviton field
contributes with $\alpha=-\tfrac{1}{16}$. Now,\begin{equation}
\kappa h^{\mathrm{q}}(t_{0})=-\frac{G}{4\pi t_{0}^{2}},\qquad\kappa\dot{h^{\mathrm{q}}}(t_{0})=\frac{G}{4\pi t_{0}^{3}}\qquad\mathrm{and}\qquad\kappa\ddot{h^{\mathrm{q}}}(t_{0})=-\frac{G}{2\pi t_{0}^{4}}.\label{eq:khq(t0) and der.}\end{equation}
The total graviton field $\kappa h=\kappa h^{\mathrm{c}}+\kappa h^{\mathrm{q}}$,
and its derivatives at the time $t_{0}$, expressed in the dimensionless
(Planck's) time unit\begin{equation}
\tau\equiv\frac{1}{\sqrt{G}}t_{0},\label{eq:def.tau}\end{equation}
are\begin{equation}
\kappa h(\tau)=-\frac{1}{4\pi\tau^{2}},\qquad\kappa\dot{h}(\tau)=-\frac{1}{t_{0}}\left(1-\frac{1}{4\pi\tau^{2}}\right)\qquad\mathrm{and}\qquad\kappa\ddot{h}(\tau)=-\frac{1}{t_{0}^{2}}\left(\frac{1}{2\pi\tau^{2}}\right).\label{eq:kh(tau) and der.}\end{equation}
Finally, by virtue of (\ref{eq:general q}), we obtain the total deceleration
parameter of the form\begin{equation}
q(\tau)=1-\frac{1}{\pi\tau^{2}}+\mathcal{O}\left(\tau^{-4}\right).\label{eq:q(tau)}\end{equation}
\begin{table}
\centering{}\begin{tabular}{|c|c|c|c|}
\hline 
spin & $\alpha_{1}$ & $\alpha_{2}$ & $\alpha$\tabularnewline
\hline
\hline 
0 & $\tfrac{1}{480}$ & $-\tfrac{1}{720}$ & 0\tabularnewline
\hline 
$\tfrac{1}{2}$ & $\tfrac{1}{160}$ & $-\tfrac{1}{240}$ & 0\tabularnewline
\hline 
1 & $\tfrac{1}{40}$ & $-\tfrac{1}{60}$ & 0\tabularnewline
\hline 
2 & $\frac{27}{80}$ & -$\frac{59}{240}$ & -$\frac{1}{16}$\tabularnewline
\hline
\end{tabular}\caption{Coefficients $\alpha_{1}$ and $\alpha_{2}$ entering the one-loop
graviton vacuum polarization (self-energy) tensor operator (\ref{eq:Pi'})
(taken from \cite{"Theone-loopneutrinocontributiontothegravitationpropagator",CalculationoftheGravitonSelf-EnergyUsingDimensionalRegularization,ComplementarityoftheMaldacenaandRandall-SundrumPictures,Onquantumcorrectionstothegravitonpropagator,Photoncorrectionstothegravitonpropagator});
$\alpha\equiv2\alpha_{1}+3\alpha_{2}$.}
\label{Flo:TABLE I}
\end{table}

\paragraph*{Final remarks.}

In the framework of the standard (one-loop) perturbative quantum gravity,
we have derived the formula (\ref{eq:q(tau)}) expressing the value
of the total (effective) deceleration parameter $q(\tau)$. The quantum
contribution, $\delta q(\tau)=q\left(\tau\right)-q^{\mathrm{c}}\left(\tau\right)\approx-\tfrac{1}{\pi\tau^{2}}$,
is negligible in our epoch, but certainly it could play a role in
a very early evolution of the Universe. Perturbative nature of the
approach imposes bounds on the applicability of the result, but nevertheless
one can observe its distinctive features: actually, it is an independent
perturbative confirmation of the existence of strong repulsive (singularity
resolving) forces typically being attributed to the realm of LQC (cosmological
bounce); inputs and outputs are consequently given in terms of the
metric tensor; only pure gravity contributes to our result (see $\alpha$
in Table \ref{Flo:TABLE I}); and finally, no trace of the UV cutoff
is present anywhere.

Nowadays, loop quantum gravity (LQG) or, in the context of cosmology,
LQC is the most promising approach towards quantization of gravity
and proper treatment of the very early evolution of the Universe.
It is interesting to compare our repulsion and the LQC bounce. In
LQC, the effective Friedmann equation is modified for extremely high
densities of the order of the so-called critical density $\rho_{\mathrm{crit}}$.
The modification amounts to the appearance of strong repulsion (bounce)
for the densities of the order of $\rho_{\mathrm{crit}}$ which are
expected in the very early Universe. The LQC bounce resembles our
repulsion. But there is some important difference between the both
approaches, besides the obvious technical differences. Namely, in
LQC the effect is due to matter fields (the density of matter $\rho$),
whereas in our approach the whole effect is exclusively due to graviton
contribution to graviton vacuum polarization (self-energy). In fact,
according to the Table \ref{Flo:TABLE I} no matter field contributes
to the repulsion.

Supported by the University of \L{}ód\'{z} grant.

\bibliographystyle{apsrev}
\bibliography{One-loop_quantum_gravity_repulsion_in_the_early_Universe2}

\end{document}